\documentclass[aps,prd,twocolumn,showpacs,superscriptaddress,nofootinbib,preprintnumbers]{revtex4-2}
\usepackage{amsmath,amssymb}
\usepackage{graphicx}
\usepackage{xcolor}
\usepackage[colorlinks=true,citecolor=blue,linkcolor=blue,urlcolor=blue]{hyperref}
\usepackage{natbib}
\usepackage{xspace}
\usepackage{aas_macros}
\usepackage{comment}
\usepackage{bbm}

\newcommand{\radxpta}{RAD$\times$PTA\xspace}

\begin{document}
\title{Optimal Statistic for a Stochastic Gravitational Wave Background using Pulsar-Timing--Astrometry Cross Correlations}

\begin{abstract}
Astrometric measurements have emerged as a viable probe of gravitational waves (GWs), complementing pulsar timing searches by providing overlapping and higher GW frequency coverage. At its core, the search for GWs using astrometry is very similar to the search with pulsar timing data: GWs induce a change in pulse arrival times along the line of sight to a pulsar and a transverse angular deflection in the apparent position of a star. In analogy to pulsar timing, we develop a correlation statistic between pairs of stellar deflections for astrometry to search for a stochastic GW background. With a common formalism for both astrometry and pulsar timing, we also develop the cross-correlation statistic between stellar deflections and pulsar timing residuals. We estimate the signal-to-noise ratios from the correlation statistics and provide scaling laws in the weak- and intermediate-signal regimes.
\end{abstract}

\author{Kimberly K.~Boddy}
\affiliation{Texas Center for Cosmology and Astroparticle Physics, Weinberg Institute for Theoretical Physics, Department of Physics, The University of Texas at Austin, Austin, TX 78712, USA}
\author{Jeffrey S. Hazboun}%ORCID[0000-0003-2742-3321]
\affiliation{Oregon State University, 1500 SW Jefferson Way, Corvallis, OR 97331}
\author{Kris Pardo}
\affiliation{University of Southern California, Los Angeles, CA 90089, USA}

\maketitle

\section{Introduction}

Multiple pulsar timing array (PTA) collaborations have reported evidence for a stochastic gravitational-wave background (GWB) at nanohertz frequencies, observed as a common red-spectrum process across pulsars whose inter-pulsar correlations follow the Hellings--Downs pattern~\cite{Hellings:1983fr} expected for an isotropic, unpolarized background~\cite{ng15gwb, eptadr2_3:gwb, pptadr3:gwb}.
The datasets and their inferred signals have been compared across collaborations~\cite{ipta3p+2024}, and the measured amplitude and spectral shape are broadly consistent with emission from a cosmological population of inspiraling supermassive black-hole binaries~\cite{burke-spolaor:2019}, though potential early Universe origins are not excluded~\cite{NANOGrav:2023hvm}.

Gravitational waves (GWs) also perturb the apparent positions of distant sources, inducing time-dependent astrometric deflections~\cite{Book:2010pf}, in close analogy to the timing residuals they produce for pulsars~\cite{Detweiler:1979wn}.
Astrometric deflections and timing residuals have different angular responses due to a GWB, and they can be combined to produce an angular cross correlation~\cite{Qin:2018yhy,Mihaylov:2018uqm}, which recent studies have started to explore through Fisher forecasts~\cite{jimenezcruz+2024, vaglio+2026}.
Together, they cover a wide range of GW frequencies: PTAs are sensitive to the lowest frequencies (${\sim} 10^{-9} - 10^{-7}~\rm{Hz}$), while high-precision astrometry and high-cadence relative astrometry detection (RAD)~\cite{Klioner:2017asb,  Wang:2020pmf, Wang:2022sxn, Pardo:2023cag} cover the microhertz (${\sim} 10^{-8} -  10^{-5}~\rm{Hz}$).
Moreover, astrometry and pulsar timing have independent systematics, further demonstrating the complementarity between these two probes.

Detecting a GWB in either data set---and in their cross-correlation---rests on measuring these angular correlations against noise.
In the pulsar-timing context, a standard tool for this purpose is the frequentist optimal cross-correlation statistic, which combines pulsar pairs with weights chosen to maximize sensitivity to the correlated signal~\cite{anholm+2009os,Chamberlin:2014ria}.
In the decade since its time-domain formulation, the statistic has been developed in several directions: marginalization over uncertain pulsar noise~\cite{vite18}; analytic null and signal distributions and associated $p$-values~\cite{hazboun+2023gx2, vanhaasteren2025a}; simultaneous fitting of multiple spatial correlations~\cite{sardesai+2023}; frequency-resolved estimation~\cite{gersbach+2025}; empirical null distributions from sky scrambles and phase shifts~\cite{cornish+2016, taylor+2017}; the covariance between distinct pulsar pairs and the associated cosmic variance~\cite{allen2022, allen+2022, allen+2025}; and posterior-predictive interpretations of the hybrid frequentist--Bayesian scheme~\cite{vallisneri+2023ppc}.
The signal-to-noise ratio of the statistic and its growth with observing time have likewise been characterized through scaling laws~\cite{siemens+2013, Vigeland:2016nmm}.
Taken together, these provide a mature framework for characterizing spatially correlated signals in timing data.

In this paper we develop the optimal cross-correlation statistic for a joint astrometry--pulsar-timing dataset.
Extending the pulsar-timing construction of Ref.~\cite{Chamberlin:2014ria} to a data vector containing both stellar deflections and pulsar timing residuals, we build the joint covariance---organized by the three overlap-reduction-function classes (RAD$\times$RAD, PTA$\times$PTA, and \radxpta)---and derive a single unbiased estimator of the background amplitude together with its decomposition into per-channel contributions.
We also indicate which of the optimal-statistic developments noted above carry over to the joint statistic.
Our work develops the detection formalism for a combined \radxpta{} analysis; numerical implementation and application to data are left to future work. Note that although we refer to RAD throughout the paper, our formalism applies to relative, differential, and absolute astrometry.\footnote{We choose RAD as our acronym, because ``stellar astrometric deflections'' just leads to a sad acronym.}

This paper is organized as follows.
We establish our conventions in Sec.~\ref{sec:conventions} in order to establish a common framework for both astrometry and PTAs.
In Sec.~\ref{sec:cov_matrix}, we build the single covariance matrix that encompasses star-star correlations for astrometry, pulsar-pulsar correlations for PTAs, and the star-pulsar cross correlations for \radxpta{}.
From the covariance matrix, we obtain the cross-correlation optimal statistic in Sec.~\ref{sec:cross-optimal} and discuss how it can be expanded to incorporate various developments of the optimal statistic for PTAs in Appendix~\ref{sec:supplements}.
Given the optimal statistic, we are able to quantify expected signal-to-noise levels and how they scale with GWB modeling parameters and specifications of the astrometric/timing surveys in Sec.~\ref{sec:scaling}.
We conclude in Sec.~\ref{sec:conclusions}.

Throughout this paper, we work in units of $c=1$.

\section{Conventions}
\label{sec:conventions}

We consider a GWB associated with the metric perturbation
\begin{equation}
  h_{ij}(t,\vec{x}) = \sum_{\alpha} \int d\Omega_k \int_{-\infty}^{\infty} df\, h_{\alpha}(f,\hat{k}) \varepsilon_{ij}^{\alpha} (\hat{k}) e^{2\pi i f (\hat{k}\cdot\vec{x} - t)} \,,
\end{equation}
where $\vec{k} = k \hat{k}$ is the wave vector associated with a given GW plane wave, $f = k/2\pi$ is the frequency, and $\alpha$ labels the two transverse traceless tensor polarization modes $\varepsilon_{ij}^\alpha (\hat{k})$, with normalization $\varepsilon_{ij}^\alpha (\hat{k}) \varepsilon^{\alpha^\prime \, ij} (\hat{k}) = 2 \delta^{\alpha\alpha^\prime}$.
We work in a coordinate system in which the origin is located at the solar system barycenter (SSB): $\vec{x}_{\rm SSB} = \vec{0}$.
For a stationary, unpolarized, statistically isotropic and homogeneous GWB, the 2-point correlation function of the Fourier amplitude $h_\alpha (f,\hat{k})$ is
\begin{equation}
  \langle h_{\alpha} (f,\hat{k}) h_{\alpha^\prime}^\ast (f',\hat{k}') \rangle = \frac{S_h(f)}{2} \delta (f-f') \frac{\delta_{\alpha\alpha^\prime}}{2} \frac{\delta^2(\hat{k}-\hat{k}')}{4\pi} \,,
\end{equation}
where $S_h(f)$ is the one-sided spectral density function, and the angle brackets denote an ensemble average.

In the remainder of this section, we present the 2-point correlation functions for the observables of interest: pulsar timing residuals and astrometric deflections.
The correlation functions separate into angular and frequency-dependent contributions, and we specify our chosen normalization for each.
We also discuss the decomposition of the astrometric deflection into orthonormal basis vectors.

\subsection{Correlation functions}

For a monochromatic GW, the timing residual of a pulsar $a$ located in the direction $\vec{n}_a$ on the celestial 2-sphere is \cite{Detweiler:1979wn}
\begin{equation}
  \delta t_a(t) = \frac{n_a^j n_a^k}{2(1+\hat{n}_a\cdot\hat{k})} \frac{1}{-2\pi i f} h_{jk}(t,\vec{0}) \,,
\end{equation}
where we have explicitly ignored the contribution from the metric evaluated at the location of the pulsar.
The components of the astrometric deflection of a star $a$ located in the direction $\vec{n}_a$ are \cite{Book:2010pf}
\begin{equation}
  \delta n_a^i = \left[ (n_a^i + k^i) \frac{n_a^j n_a^k}{2 (1+\hat{n}_a\cdot\hat{k})} - \frac{1}{2} n_a^j \delta^{ik} \right] h_{jk}(t,\vec{0}) \,,
\end{equation}
where we keep only the leading-order terms in the limit where the distance to the star is large compared to the GW wavelength.

For a background of GWs, the timing residual and astrometric deflection become
\begin{align}
  \delta t_a(t) &= \sum_\alpha\!\int \!d\Omega_k\!\!\int_{-\infty}^{\infty}\!\!\!df\;
  h_\alpha(f,\hat k)\, \frac{F_a^{\alpha}(\hat k)}{-2\pi if}\,e^{-2\pi i f t} \\
  \delta n^i_a(t) &= \sum_\alpha\!\int \!d\Omega_k\!\!\int_{-\infty}^{\infty}\!\!\!df\;
  h_\alpha(f,\hat k)\,F_a^{\alpha,i}(\hat k)\,e^{-2\pi i f t} \, ,
\end{align}
where their respective detector response functions are
\begin{align}
  F_a^{\alpha}(\hat{k}) &\equiv \frac{n_a^j n_a^k}{2(1+\hat{n}_a\cdot\hat{k})} \varepsilon_{jk}^{\alpha} (\hat{k}) \\
  F_a^{\alpha, i}(\hat{k}) &\equiv \left[ (n_a^i + k^i) \frac{n_a^j n_a^k}{2 (1+\hat{n}_a\cdot\hat{k})} - \frac{1}{2} n_a^j \delta^{ik} \right] \varepsilon_{jk}^{\alpha} (\hat{k}) \, .
\end{align}

Since we are interested in the various 2-point correlation functions that can be formed with these two observables, we define the unnormalized overlap reduction functions (ORFs)
\begin{align}
  \Gamma_{ab}^{pp} &\equiv \frac{1}{8\pi} (1+\delta_{ab}) \sum_\alpha \int d\Omega_k\, F^\alpha_a(\hat{k}) \left[F^{\alpha}_b(\hat{k})\right]^\ast \\
  \Gamma_{ab}^{p\star, i} &\equiv \frac{1}{8\pi} \sum_\alpha \int d\Omega_k\, F^\alpha_a(\hat{k}) \left[F^{\alpha, i}_b(\hat{k})\right]^\ast \\
  \Gamma_{ab}^{\star\star, ij} &\equiv \frac{1}{8\pi} \sum_\alpha \int d\Omega_k\, F^{\alpha,i}_a(\hat{k}) \left[F^{\alpha, j}_b(\hat{k})\right]^\ast \, ,
\end{align}
denoting PTA pulsars as $p$ and RAD stars as $\star$ to label the ORFs.
The PTA$\times$PTA ORF includes an additional $\delta_{ab}$ to account for the autocorrelation contribution.

In line with conventions in the PTA community, we normalize the PTA-PTA and the RAD-RAD autocorrelation ORFs to be unity (for the great-arc basis presented in Sec.~\ref{sec:deflection-decomp}).
The normalized ORFs and the resulting normalized cross-correlation ORF are
\begin{align}
  \xi_{ab}^{pp} &= 3\, \Gamma_{ab}^{pp} \\
  \xi_{ab}^{p\star,i} &= 6\, \Gamma_{ab}^{p\star, i} \\
  \xi_{ab}^{\star\star,ij} &= 12 \, \Gamma_{ab}^{\star\star,ij} \, .
\end{align}
Defining the power spectral densities
\begin{align}
  \mathcal{P}_\mathrm{GW}^{pp}(f) &= \frac{1}{12 \pi^2 f^2} S_h(f) \\
  \mathcal{P}_\mathrm{GW}^{p\star}(f) &= \frac{1}{12 \pi f} S_h(f) \\
  \mathcal{P}_\mathrm{GW}^{\star\star}(f) &= \frac{1}{12} S_h(f) \, ,
\end{align}
the 2-point correlation functions can be written as
\begin{widetext}
\begin{align}
  \langle \delta t_a(t) \delta t_b(t') \rangle
  &= \int_0^\infty df\, \frac{S_h(f)}{(2\pi f)^2} \cos(2\pi f\tau) \Gamma_{ab}^{pp}
  = \int_0^\infty df\, \mathcal{P}_\mathrm{GW}^{pp}(f)\, \cos(2\pi f\tau)\, \xi_{ab}^{pp} \\
  \langle \delta t_a(t) \delta n_b^i(t') \rangle
  &= \int_0^\infty df\, \frac{S_h(f)}{2\pi f} \sin(2\pi f\tau) \Gamma_{ab}^{p\star,i}
  = \int_0^\infty df\, \mathcal{P}_\mathrm{GW}^{p\star}(f)\, \sin(2\pi f\tau)\, \xi_{ab}^{p\star,i} \label{eqn:cross-2pt-func} \\
  \langle \delta n_a^i(t) \delta n_b^j(t') \rangle
  &= \int_0^\infty df\, S_h(f) \cos(2\pi f\tau) \Gamma_{ab}^{\star\star,ij}
  = \int_0^\infty df\, \mathcal{P}_\mathrm{GW}^{\star\star}(f)\, \cos(2\pi f\tau)\, \xi_{ab}^{\star\star,ij} \, ,
\end{align}
\end{widetext}
where $\tau \equiv t-t'$.

Note that the \radxpta cross correlation has a sine rather than a cosine dependence, since $S_h(f)$ is even in $f$, while the factor of $S_h(f)/(2\pi f)$ is odd.
Thus, the deflections, measured in strain, are $\pi/2$ out of phase with the pulsar timing residuals.
This is known from Cram\'er's spectral representation for multivariate stationary processes\footnote{In particular, see Ch.~9 (multivariate spectra; co-spectrum and quadrature spectrum) and the Hermitian nonnegative-definite spectral density matrix of Ref.~\cite{priestley1981}.} \cite{cramer1940,priestley1981}.
Although Eq.~\eqref{eqn:cross-2pt-func} is antisymmetric in $\tau\equiv t-t'$ because of the $\sin[2\pi f (t-t')]$ in the integral, it does not change the symmetry of the full covariance matrix, as we clarify in Section~\ref{sec:cov_matrix}.

If we assume the spectral density function takes a power-law form
\begin{equation}
    S_h(f) = A_{\rm gw}^2 \frac{1}{f} \left(\frac{f}{f_0}\right)^{2\alpha} ,
\end{equation}
which arises from writing the characteristic strain as $h_c(f)=A_{\rm gw}(f/f_0)^{\alpha}$, with $S_h=h_c^2/f$.
The power spectral densities become
\begin{align}
  \mathcal{P}^{pp}_{\rm GW}(f) &= \frac{A_{\rm gw}^2}{12\pi^2 f_0^{2\alpha}}\, f^{-(3-2\alpha)}
     \equiv \mathcal{A}_p\,f^{-\beta} \; , \label{eq:P_pp} \\
  \mathcal{P}^{\star\star}_{\rm GW}(f) &= \frac{A_{\rm gw}^2}{12\, f_0^{2\alpha}}\, f^{-(1-2\alpha)}
     \equiv \mathcal{A}_\star\,f^{-\gamma} \; , \label{eq:P_ss} \\
  \mathcal{P}^{p\star}_{\rm GW}(f) &= \frac{A_{\rm gw}^2}{12\pi f_0^{2\alpha}} f^{-(2-2\alpha)}
     = \sqrt{\mathcal{A}_p\mathcal{A}_\star}\; f^{-(\beta+\gamma)/2} \; , \label{eq:P_ps}
\end{align}
with $\beta = 3-2\alpha$ and $\gamma = 1-2\alpha$.

\subsection{Decomposition of astrometric deflection}
\label{sec:deflection-decomp}

\begin{figure}[t]
    \centering
    \includegraphics[width=\linewidth]{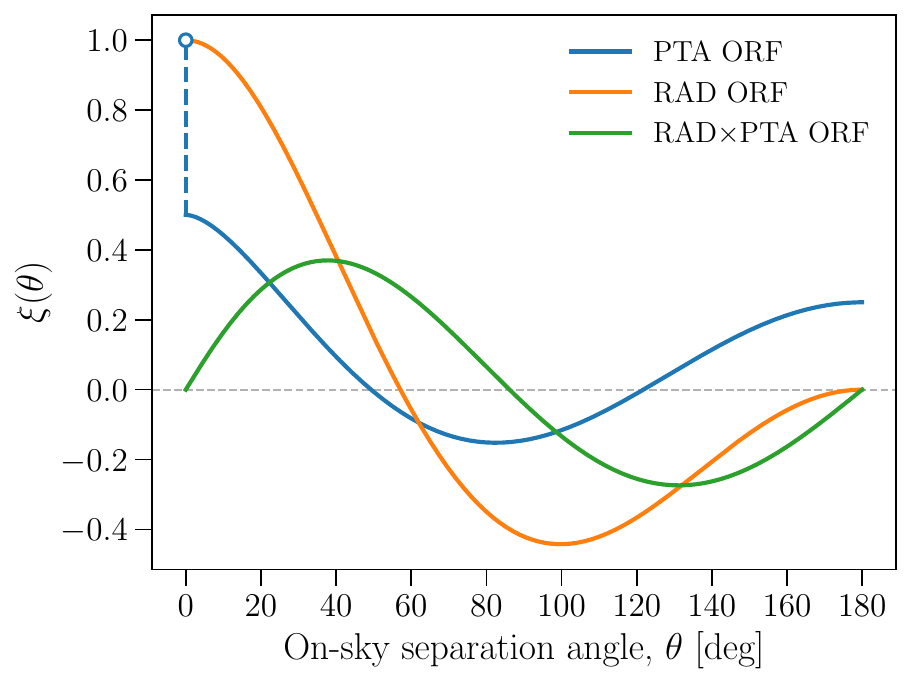}
    \caption{Normalized overlap reduction functions (ORFs) for PTA$\times$PTA (blue), RAD$\times$RAD (orange), and \radxpta (green).
      The PTA$\times$PTA and RAD$\times$RAD curves are normalized such that their autocorrelation is unity.
      Note that the PTA$\times$PTA curve can take two values at zero separation angle.
      The normalized autocorrelation of a pulsar is $1$, but the correlation between pulsars at the same sky location and different distances is $1/2$.}
    \label{fig:orfs}
\end{figure}

The 2-point correlation functions involving astrometric deflections depend on the basis chosen to decompose the deflection vector on the 2-sphere.
For a given star located in the direction $\vec{n}_a$, a convenient basis is defined with respect to the great arc connecting the star to another star or pulsar located in the direction $\vec{n}_b$:
\begin{equation}
    \delta \vec{n}_a \;=\; \delta n^\parallel_a \,\hat{e}^\parallel_{ab} \;+\; \delta n^\perp_a \,\hat{e}^\perp_{ab} \; ,
\end{equation}
with the unit vectors~\cite{Mihaylov:2018uqm}
\begin{equation}
    \hat{e}^\parallel_{ab} = \frac{(\hat{n}_a \times \hat{n}_b) \times \hat{n}_a}{1 - (\hat{n}_a \cdot \hat{n}_b)^2}
    \; , \quad
    \hat{e}^\perp_{ab} = \frac{\hat{n}_a \times \hat{n}_b}{1 - (\hat{n}_a \cdot \hat{n}_b)^2}
\end{equation}
that point parallel and perpendicular to the great arc, respectively.
We may also recast the astrometric detector response function in this basis in order to write
\begin{align}
    \Gamma_{ab}^{p\star, i} &= \Gamma_{ab}^{p\star, \parallel} (\hat{e}^\parallel_{ab})^i + \Gamma_{ab}^{p\star, \perp} (\hat{e}^\perp_{ab})^i \\
    \Gamma_{ab}^{\star\star,ij} &= \Gamma_{ab}^{\star\star, \parallel\parallel} (\hat{e}^\parallel_{ab})^i (\hat{e}^\parallel_{ab})^j + \Gamma_{ab}^{\star\star, \perp\perp} (\hat{e}^\perp_{ab})^i (\hat{e}^\perp_{ab})^j \nonumber \\
    &\quad + \Gamma_{ab}^{\star\star, \parallel\perp} (\hat{e}^\parallel_{ab})^i (\hat{e}^\perp_{ab})^j + \Gamma_{ab}^{\star\star, \perp\parallel} (\hat{e}^\perp_{ab})^i (\hat{e}^\parallel_{ab})^j \, .
\end{align}
In the absence of explicit parity breaking, $\Gamma_{ab}^{\star\star, \parallel\perp} = \Gamma_{ab}^{\star\star, \perp\parallel} = \Gamma_{ab}^{p\star, \perp} = 0$; moreover, the surviving RAD$\times$RAD ORF components are equal~\cite{Qin:2018yhy,Mihaylov:2018uqm}.
The normalized ORFs, as shown in Fig.~\ref{fig:orfs}, become
\begin{align}
  \xi_{ab}^{pp} &= \frac{1+\delta_{ab}}{2} - \frac{x_{ab}^2}{4} + 3x_{ab}^2 \ln(x_{ab}) \\
  \xi_{ab}^{p\star,\parallel} &= x_{ab}(1-x_{ab}^2) + \frac{3x_{ab}^3}{1-x_{ab}^2} \ln(x_{ab}) \\
  \xi_{ab}^{\star\star,\parallel\parallel} = \xi_{ab}^{\star\star,\perp\perp} &= 1 - 7 x_{ab}^2 - \frac{12 x_{ab}^4}{1-x_{ab}^2} \ln(x_{ab}) \, ,
\end{align}
where $x_{ab} \equiv \sin(\theta_{ab} / 2)$ and $\theta_{ab} \equiv \cos^{-1} (\hat{n}_a \cdot \hat{n}_b)$ is the angle between the star/pulsar pairs.

We caution that the theoretical curves for RAD$\times$RAD and \radxpta in Fig.~\ref{fig:orfs} do not immediately connect to data, because we have chosen a coordinate system that differs for each pair of objects.
A global coordinate system, such as Galactic coordinates, is more appropriate in order to construct a unified covariance matrix; the theoretical ORFs must be transformed from the pairwise great-arc basis to a global basis.

\section{Covariance Matrices}
\label{sec:cov_matrix}

In order to discuss the optimal statistic, we must first construct the covariance matrices for each of our cases.
We begin by reviewing the PTA covariance matrix construction. We then construct the astrometry-only and the cross-correlation covariance matrices.

\subsection{Pulsar Timing Array Covariance Matrix}

For the pulsar timing covariance matrix, we follow Ref.~\cite{Chamberlin:2014ria}.
The GW signal appears in the pulsar timing residuals, which are the differences between the measured pulse arrival times and a fitted timing model for the pulsar.
The PTA covariance has auto and cross terms.
For a pulsar $a$ observed at $N_a$ epochs, let $\delta\mathbf{t}_a$ be the
length-$N_a$ vector of timing residuals. Its auto-covariance is the
$N_a\times N_a$ matrix
\begin{equation}
    \tilde{P}_a = \langle \delta\mathbf{t}_a\, \delta\mathbf{t}_a^{T}\rangle \; ,
\end{equation}
and the cross-covariance between pulsars $a$ and $b$ is the $N_a\times N_b$ matrix
\begin{equation}
    \tilde{T}_{ab} = \langle \delta\mathbf{t}_a\, \delta\mathbf{t}_b^{T} \rangle \; ,
    \qquad a \neq b \; .
\end{equation}
In the frequency domain, the diagonal of $\tilde{P}_a$ is set by the per-pulsar
spectral density $\mathcal{P}^{p}_a(f) = \mathcal{P}^{p}_{a,\mathrm{WN}}
+ \mathcal{P}^{p}_{a,\mathrm{IRN}} + \mathcal{P}^{pp}_{\mathrm{GW}}$ (white noise,
intrinsic red noise, and the GWB).
Throughout, boldface denotes a data or residual vector, and plain capital Latin letters denote matrices.

Before these enter a detection statistic, the fit to the timing model must be
propagated into the covariance. For pulsar $a$ with $N_a$ arrival times, let $M_a$
be the $N_a \times N^{\rm param}_a$ timing-model design matrix, whose columns are the
derivatives of the arrival times with respect to the $N^{\rm param}_a$ timing-model
parameters. These fitted parameters are marginalized over by projecting onto the
complement of the design matrix: we introduce $G_a$, an
$N_a \times (N_a - N^{\rm param}_a)$ matrix whose columns form an orthonormal basis for the
null space of $M_a^{T}$, so that~\cite{vanHaasterenLevin2013, Chamberlin:2014ria}
\begin{equation}
    G_a^{T} M_a = 0 \; , \qquad G_a^{T} G_a = \mathbbm{1} \; .
    \label{eq:Gdef}
\end{equation}
The marginalized residuals are the inner product of the raw residuals with $G_a$,
\begin{equation}
    \mathbf{r}_a \equiv G_a^{T}\, \delta\mathbf{t}_a \; ,
\end{equation}
and every covariance block is reduced by the same inner product,
\begin{equation}
    P_a = G_a^{T}\, \tilde{P}_a\, G_a \; , \qquad
    T_{ab} = G_a^{T}\, \tilde{T}_{ab}\, G_b \; .
    \label{eq:pta-reduced}
\end{equation}
For a set of $N_p$ pulsars, the full covariance in this reduced basis is
\begin{equation}
    C =  \begin{pmatrix}
        P_1 & T_{12} & \cdots & T_{1N_p} \\
        T_{12}^{T} & P_2 & & \vdots \\
        \vdots & & \ddots & \\
        T_{1N_p}^{T} & \cdots & & P_{N_p}
    \end{pmatrix} \; .
\end{equation}
Stacking the per-pulsar reduced residuals into the single length-$\sum_a(N_a-N^{\rm param}_a)$ vector
$\mathbf{r} = (\mathbf{r}_1, \mathbf{r}_2, \ldots, \mathbf{r}_{N_p})$, the Gaussian
likelihood is
\begin{equation}
    \ln\mathcal{L} \sim -\tfrac{1}{2}\, \mathbf{r}^{T} C^{-1} \mathbf{r}
    - \tfrac{1}{2}\ln\det(2\pi C) \; ,
\end{equation}
from which the optimal statistic is constructed below.

\subsection{Astrometry Covariance Matrix}

We decompose a star's angular deflection on the sky $\delta \vec{n}$ into two orthogonal components
\begin{equation}
    \delta \vec{n} \;=\; \delta x \,\hat{x} \;+\; \delta y \,\hat{y} \; ,
\end{equation}
where $\hat{x}$ and $\hat{y}$ represent the unit basis vectors in a global coordinate system (cf.\ Sec.~\ref{sec:deflection-decomp}).
Each star thus contributes two data streams, and the covariance is built from blocks for each pair of components.
The auto-covariance noise power of a single star~$a$ in either component is
\begin{align}
    \widetilde{P}^{\,x}_{a}
    &= \bigl\langle \delta\mathbf{x}_a\; \delta\mathbf{x}_a^{T} \bigr\rangle = \int_0^\infty df\, \mathcal{P}_a^{\star\,x}(f)\, \cos(2\pi f\tau) \\
    \widetilde{P}^{\,y}_{a}
    &= \bigl\langle \delta\mathbf{y}_a\; \delta\mathbf{y}_a^{T} \bigr\rangle = \int_0^\infty df\, \mathcal{P}_a^{\star\,y}(f)\, \cos(2\pi f\tau) \; ,
\end{align}
where the spectral density is the sum of white noise, intrinsic red noise,
and a GWB,
\begin{equation}
    \mathcal{P}_a^{\star}(f) \;=\; \mathcal{P}^{\star}_{a,\mathrm{WN}}
    \;+\; \mathcal{P}^{\star}_{a,\mathrm{IRN}}
    \;+\; \mathcal{P}_\mathrm{GW}^{\star} \; .
\end{equation}
The cross-star correlations between stars $a$ and $b$ are
\begin{align}
    \widetilde{X}_{ab}  &\;=\; \bigl\langle \delta\mathbf{x}_a\; \delta\mathbf{x}_b^{T} \bigr\rangle \;, &
    \widetilde{Y}_{ab}  &\;=\; \bigl\langle \delta\mathbf{y}_a\; \delta\mathbf{y}_b^{T} \bigr\rangle \;, \\
    \widetilde{V}_{ab}  &\;=\; \bigl\langle \delta\mathbf{x}_a\; \delta\mathbf{y}_b^{T} \bigr\rangle \;, &
    \widetilde{U}_{ab}  &\;=\; \bigl\langle \delta\mathbf{y}_a\; \delta\mathbf{x}_b^{T} \bigr\rangle \;, \\
    \widetilde{W}_{a}   &\;=\; \bigl\langle \delta\mathbf{x}_a\; \delta\mathbf{y}_a^{T} \bigr\rangle \;.
\end{align}

The deterministic-model marginalization proceeds exactly as in the pulsar-timing case [Eqs.~\eqref{eq:Gdef}--\eqref{eq:pta-reduced}]: each component is projected onto the complement of a design matrix $G_a$ through the same inner product.
What differs is the content of the model---for astrometry, $G_a$ removes not a source astrometric model, but the spacecraft systematics and differential velocity aberration \cite[for some example systematics, see, e.g.,][]{Monet:2010, VanCleve:2016kdch, VanCleve:2016kih}.
The reduced residuals are
\begin{equation}
    \mathbf{r}^{x}_{a} = G_a^{T}\,\delta\mathbf{x}_a \;, \qquad
    \mathbf{r}^{y}_{a} = G_a^{T}\,\delta\mathbf{y}_a \;, \qquad
    \mathbf{r}^{\star}_{a} = \begin{pmatrix}\mathbf{r}^{x}_{a}\\[2pt] \mathbf{r}^{y}_{a}\end{pmatrix} \;,
\end{equation}
and the reduced blocks are
\begin{equation}
    P^{x,y}_{a} = G_a^{T}\,\widetilde{P}^{\,x,y}_{a}\,G_a \;, \qquad
    X_{ab} = G_a^{T}\,\widetilde{X}_{ab}\,G_b \;,
\end{equation}
and likewise for $Y_{ab}$, $U_{ab}$, $V_{ab}$, and $W_a$. Stacking the $x$ and $y$
streams, the reduced covariance is
\begin{equation}
    C_{ab} \;=\;
    \begin{pmatrix}
        P^{x}_{a}   & W_{a}     & X_{ab}    & V_{ab}    & \cdots \\[0.25em]
        W^{T}_{a}   & P^{y}_{a} & U_{ab}    & Y_{ab}    &        \\[0.25em]
        X^{T}_{ab}  & U^{T}_{ab}& P^{x}_{b} & W_{b}     &        \\[0.25em]
        V^{T}_{ab}  & Y^{T}_{ab}& W^{T}_{b} & P^{y}_{b} &        \\[0.25em]
        \vdots      &           &           &           & \ddots
    \end{pmatrix} \; .
\end{equation}
Although the pairwise great-arc basis of Sec.~\ref{sec:deflection-decomp} sets $U_{ab}=V_{ab}=0$ for any single pair, no global basis achieves this for all pairs simultaneously, so we retain the full covariance $C_{ab}$ with all component blocks.

\subsection{Cross-Correlation Covariance Matrix}

To cross-correlate astrometry and pulsar timing we stack the reduced residuals
defined above into a single data vector $\mathbf{s}$, given by
\begin{equation}
  \mathbf{s} =
  \begin{cases}
    \;\mathbf{r}^{\star} \\
    \;\mathbf{r} 
  \end{cases}
\end{equation}
stacked stars first and pulsars second. We work throughout in the reduced basis,
so the raw residuals and the design-matrix projections of the previous two
subsections are left implicit from here on. The covariance of $\mathbf{s}$ is
built from four block types,
\begin{align}
P_a    &= \langle \mathbf{s}_a\, \mathbf{s}_a^{T} \rangle & &\text{(auto, star or pulsar)} \\
Z_{ab} &= \langle \mathbf{s}_a\, \mathbf{s}_b^{T} \rangle,\ a \neq b & &\text{(star $\times$ star)} \\
T_{ab} &= \langle \mathbf{s}_a\, \mathbf{s}_b^{T} \rangle,\ a \neq b & &\text{(pulsar $\times$ pulsar)} \\
Q_{ab} &= \langle \mathbf{s}_a\, \mathbf{s}_b^{T} \rangle & &\text{(pulsar $\times$ star).}
\end{align}
A single index pair $a,b$ runs over all $N_\star+N_p$ objects; the block type---and the
star/pulsar superscript on the auto-term---records whether each index labels a
star or a pulsar. We use this same convention, $a,b$ as indices of the one
combined covariance, throughout the construction of the cross statistic below.
For $N_\star$ stars and $N_p$ pulsars the full covariance reads
\begin{equation}
\Sigma \!=\!\!\! \begin{pmatrix}
P^\star_a        & Z_{ab}     & \cdots & Z_{aN_\star}     & Q_{ca}^{T} & Q_{da}^{T} & \cdots & Q_{N_pa}^{T} \\
Z_{ab}^{T}       & P^\star_b  & \cdots & Z_{bN_\star}     & Q_{cb}^{T} & Q_{db}^{T} & \cdots & Q_{N_pb}^{T} \\
\vdots           & \vdots     & \ddots & \vdots           & \vdots     & \vdots     &        & \vdots       \\
Z_{aN_\star}^{T} & Z_{bN_\star}^{T} & \cdots & P^{\star}_{N_\star} & Q_{cN_\star}^{T} & Q_{dN_\star}^{T} & \cdots & Q_{N_pN_\star}^{T} \\
Q_{ca}           & Q_{cb}     & \cdots & Q_{cN_\star}     & P^p_c      & T_{cd}     & \cdots & T_{cN_p}     \\
Q_{da}           & Q_{db}     & \cdots & Q_{dN_\star}     & T_{cd}^{T} & P^p_d      & \cdots & T_{dN_p}     \\
\vdots           & \vdots     &        & \vdots           & \vdots     & \vdots     & \ddots & \vdots       \\
Q_{N_pa}         & Q_{N_pb}   & \cdots & Q_{N_pN_\star}   & T_{cN_p}^{T} & T_{dN_p}^{T} & \cdots & P^p_{N_p}
\end{pmatrix}\nonumber
\end{equation}
with $P^\star_a$ and $P^p_a$ the star and pulsar auto-terms.

Unlike the astrometry-only case, $\Sigma$ does not block-simplify under a coordinate choice: a GWB induces every block type ($P$, $Z$, $T$, and $Q$).
The star--pulsar blocks carry the antisymmetric ($\sin$) kernel of Eq.~\eqref{eqn:cross-2pt-func}, yet $\Sigma$ remains symmetric, because each $Q_{ab}$ appears with its transpose in the mirror block.

\section{Cross-Correlation Optimal Statistic}
\label{sec:cross-optimal}

With the full covariance matrix, we are able to formulate a likelihood for the data vector $\mathbf{s}$, given the presence of a GWB with an amplitude $A_{\rm gw}$.
We then use this likelihood to construct the optimal statistic for the amplitude, which allows us to define a signal-to-noise ratio.
The construction of our optimal statistic follows standard PTA literature, and in Appendix~\ref{sec:supplements}, we discuss how various developments of the optimal statistic within the PTA community apply to our case.

The full covariance matrix separates into a block-diagonal piece $P_a$ carrying
the autocorrelations, which absorbs the individual intrinsic star/pulsar noise
and the auto contribution of the GWB, and an off-diagonal piece $S_{ab}$ that
accounts for the spatial cross correlations,
\begin{equation}
    \Sigma_{ab} \;=\; P_a\,\delta_{ab} \;+\; S_{ab},\qquad S_{aa}=0 \; .
    \label{eq:auto-cross-split}
\end{equation}
We factor the (squared) GWB amplitude out of the correlated block,
\begin{equation}
    S_{ab} \;=\; A_{\rm gw}^{2}\,\bar{S}_{ab} \; ,
    \label{eq:Sbar-def}
\end{equation}
so that $\bar{S}_{ab}$ carries information about only the spectral shape and the ORF.
The correlated block takes three forms, depending on what kind of object $a$ and $b$ are: star--star ($\bar{S}^{\star\star}_{ab}\equiv Z_{ab}/A_{\rm gw}^{2}$), pulsar--pulsar ($\bar{S}^{pp}_{ab}\equiv T_{ab}/A_{\rm gw}^{2}$), and star--pulsar ($\bar{S}^{\star p}_{ab}\equiv Q_{ab}^{T}/A_{\rm gw}^{2}$).
Each carries the GWB spectral shape together with the appropriate ORF: Hellings--Downs for pulsar pairs, the astrometric ORFs for star pairs, and the mixed pulsar--astrometry ORF for the cross pairs.

Working in the weak-signal regime $A_{\rm gw}^{2}\bar{S}\ll P$, a first-order expansion of the Gaussian log-likelihood gives, up to $\mathbf{s}$-independent terms,
\begin{align}
    \ln p(\mathbf{s}|A_{\rm gw}^{2}) \;\approx &\; -\tfrac{1}{2}\sum_{a}\mathbf{s}_a^{T}P_a^{-1}\mathbf{s}_a \nonumber \\
    &+ A_{\rm gw}^{2}\!\sum_{a<b}\mathbf{s}_a^{T}P_a^{-1}\bar{S}_{ab}P_b^{-1}\mathbf{s}_b,
    \label{eq:loglike-expansion}
\end{align}
where the sum for $a<b$ runs over all \emph{unique} pairings of $a$ and $b$, which are drawn from all $N_\star+N_p$ stars and pulsars combined. Note that the $\ln\det\Sigma$ contribution vanishes at this order because $\bar{S}$ is block off-diagonal. The log-likelihood ratio between the correlated-signal and noise-only hypotheses is therefore
\begin{equation}
    \ln\Lambda \equiv \ln \frac{p(\mathbf{s}|A_{\rm gw}^2)}{p(\mathbf{s}|0)} \approx A_{\rm gw}^{2}\,\hat N ,
    \label{eq:lnL}
\end{equation}
where
\begin{equation}
    \hat N \;\equiv\; \sum_{a<b}\mathbf{s}_a^{T}P_a^{-1}\bar{S}_{ab}P_b^{-1}\mathbf{s}_b ,
\end{equation}
and the hat on the $N$ designates it as an estimator. The linear functional $\hat N$ has expectation value $\langle\hat N\rangle = A_{\rm gw}^{2}\mathcal{N}$, where
\begin{equation}
    \mathcal{N} \;\equiv\; \sum_{a<b}\mathrm{tr}\!\left[P_a^{-1}\bar{S}_{ab}P_b^{-1}\bar{S}_{ba}\right].
    \label{eq:Ndef}
\end{equation}

Demanding an unbiased frequentist estimator of the squared amplitude fixes the normalization ($\bigl\langle\hat A^{2}_{\rm joint}\bigr\rangle = A_{\rm gw}^{2}$) and yields a single joint \radxpta optimal statistic
\begin{equation}
    \;\hat A^{2}_{\rm joint} \;=\; \frac{\hat N}{\mathcal{N}} \; .
    \label{eq:A2-joint-v2}
\end{equation}
Under the null hypothesis of no GWB, $\mathrm{Var}(\hat N)=\mathcal{N}$ to leading order, so the null standard deviation of $\hat A^{2}_{\rm joint}$ is $\sigma_{0}=\mathcal{N}^{-1/2}$, and the signal-to-noise ratio (SNR) is
\begin{equation}
    \hat{\rho}_{\rm joint} \;=\; \frac{\hat A^{2}_{\rm joint}}{\sigma_{0}} \;=\; \frac{\hat N}{\sqrt{\mathcal{N}}}
    \label{eq:rho-joint-v2}
\end{equation}
with an expectation value $\langle\hat{\rho}_{\rm joint}\rangle = A_{\rm gw}^{2}\sqrt{\mathcal{N}}$.
Equations~\eqref{eq:A2-joint-v2} \& \eqref{eq:rho-joint-v2} define a \emph{single} estimator with a \emph{single} normalization $\mathcal{N}$; the normalization $\mathcal{N}$ and the variance $\sigma_0^{2}=1/\mathcal{N}$ are properties of the full dataset and are \emph{not} separable into pair classes at this level.
 
\subsection{Natural decomposition at the log-likelihood-ratio level}
 
The pair sum in Eq.~\eqref{eq:Ndef} partitions into three disjoint classes,
\begin{equation}
    \sum_{a<b} \;=\; \underbrace{\sum_{\substack{a<b\\\mathrm{stars}}}}_{\rm RAD} \;+\; \underbrace{\sum_{\substack{a<b\\\mathrm{pulsars}}}}_{\rm PTA} \;+\; \underbrace{\sum_{\substack{a\in\mathrm{stars}\\b\in\mathrm{pulsars}}}}_{\rm RAD\times PTA},
    \label{eq:pair-partition}
\end{equation}
so both $\hat N$ and $\mathcal{N}$ split linearly,
\begin{align}
    \hat N \;=&\; \hat N_{\rm RAD}+\hat N_{\rm PTA}+\hat N_{\rm RAD\times PTA} \nonumber \\
    \mathcal{N} \;=&\; \mathcal{N}_{\rm RAD}+\mathcal{N}_{\rm PTA}+\mathcal{N}_{\rm RAD\times PTA}.
    \label{eq:split-numerator}
\end{align}
Substituting into Eq.~\eqref{eq:lnL}, the log-likelihood ratio decomposes additively,
\begin{equation}
    \ln\Lambda \;\approx\; A_{\rm gw}^{2}\bigl(\hat N_{\rm RAD}+\hat N_{\rm PTA}+\hat N_{\rm RAD\times PTA}\bigr).
    \label{eq:lnL-split}
\end{equation}
Equation~\eqref{eq:lnL-split} is the \emph{natural} place at which the three pair classes separate: each contribution is an independent linear functional of the data, tied to a distinct angular ORF. The joint estimator $\hat A^{2}_{\rm joint}$ itself does \emph{not} decompose into three per-class estimators, because its denominator $\mathcal{N}$ is fixed by unbiasedness when summed over \emph{all} pairs; the ratios $\hat N_c/\mathcal{N}_c$ for $c \in \{ {\rm RAD}, {\rm PTA}, {\rm RAD\times PTA} \}$ do not individually estimate $A_{\rm gw}^{2}$.
 
\subsection{Quadrature-sum SNR at the expectation level}
 
At the level of the expected SNR, the decomposition does carry through cleanly, precisely because $\mathcal{N}$ is linear in the pair sum. Defining per-class contributions to the squared SNR by $\langle\hat{\rho}_{c}\rangle^{2}\equiv A_{\rm gw}^{4}\,\mathcal{N}_{c}$, Eq.~\eqref{eq:rho-joint-v2} yields
\begin{equation}
    \langle\hat{\rho}_{\rm joint}\rangle^{2} \!=\! A_{\rm gw}^{4}\,\mathcal{N} = \langle\hat{\rho}_{\scriptscriptstyle\rm RAD}\rangle^{2}+\langle\hat{\rho}_{\scriptscriptstyle \rm PTA}\rangle^{2}+\langle\hat{\rho}_{\scriptscriptstyle\rm RAD\times PTA}\rangle^{2}.
    \label{eq:rho-quadsum}
\end{equation}
The PTA contribution
\begin{equation}
    \langle\hat{\rho}_{\rm PTA}\rangle^{2} \;=\; A_{\rm gw}^{4}\!\!\sum_{\substack{a<b\\\mathrm{pulsars}}}\!\mathrm{tr}\!\left[(P_a^{p})^{-1}\bar{S}^{pp}_{ab}(P_b^{p})^{-1}\bar{S}^{pp}_{ba}\right],
    \label{eq:rho-PTA-v2}
\end{equation}
exactly matches Ref.~\cite{Chamberlin:2014ria}, while the RAD contribution takes the same form over the two-component star blocks,
\begin{equation}
    \langle\hat{\rho}_{\rm RAD}\rangle^{2} \;=\; A_{\rm gw}^{4}\!\!\sum_{\substack{a<b\\\mathrm{stars}}}\!\mathrm{tr}\!\left[(P^{\star}_a)^{-1}\bar{S}^{\star\star}_{ab}(P^{\star}_b)^{-1}\bar{S}^{\star\star}_{ba}\right],
    \label{eq:rho-RAD-v2}
\end{equation}
and the cross contribution sums over all $N_\star\!\cdot\! N_p$ star--pulsar pairs,
\begin{equation}
    \langle\hat{\rho}_{\rm RAD\times PTA}\rangle^{2} \;=\; A_{\rm gw}^{4}\!\sum_{\substack{a\in\mathrm{stars}\\b\in\mathrm{pulsars}}}\!\mathrm{tr}\!\left[(P_a^{\star})^{-1}\bar{S}^{\star p}_{ab}(P_b^{p})^{-1}\bigl(\bar{S}^{\star p}_{ab}\bigr)^{T}\right].
    \label{eq:rho-cross-v2}
\end{equation}
Because the dominant noise processes for stars and pulsars are largely independent, the cross-channel contribution is clean under the null hypothesis, and shared systematics (clock errors, reference-frame or attitude correction residuals) enter with angular patterns distinct from a true GWB---making the cross correlation a useful systematics diagnostic as well as a detection channel.
 
\subsection{Per-channel sub-optimal estimators}
 
If one wishes to analyze the three pair classes independently, one can define three separate unbiased estimators,
\begin{equation}
    \hat A^{2}_{c} \;\equiv\; \hat N_{c}/\mathcal{N}_{c},\qquad c\in\{\mathrm{\scriptstyle RAD},\,\mathrm{\scriptstyle PTA},\,\mathrm{\scriptstyle RAD\times PTA}\},
    \label{eq:per-channel-estimators}
\end{equation}
each satisfying $\langle\hat A^{2}_{c}\rangle=A_{\rm gw}^{2}$ with null variance $1/\mathcal{N}_{c}$. These are \emph{not} a decomposition of $\hat A^{2}_{\rm joint}$; rather, the joint estimator is their inverse-variance weighted combination,
\begin{equation}
    \hat A^{2}_{\rm joint} \!=\! \frac{\mathcal{N}_{\scriptscriptstyle\rm RAD}\,\hat A^{2}_{\scriptscriptstyle\rm RAD}\!+\!\mathcal{N}_{\scriptscriptstyle\rm PTA}\,\hat A^{2}_{\scriptscriptstyle\rm PTA}\!+\!\mathcal{N}_{\scriptscriptstyle\rm RAD\times PTA}\,\hat A^{2}_{\scriptscriptstyle\rm RAD\times PTA}}{\mathcal{N}_{\scriptscriptstyle\rm RAD}+\mathcal{N}_{\scriptscriptstyle\rm PTA}+\mathcal{N}_{\scriptscriptstyle\rm RAD\times PTA}}.
    \label{eq:ivw}
\end{equation}
In the case of only using astrometric data, $\hat A^{2}_{\rm RAD}$ is itself the
optimal statistic for a stand-alone astrometry analysis; written out,
\begin{equation}
    \hat A^{2}_{\rm RAD} \;=\;
    \frac{\displaystyle\sum_{a<b}
        (\mathbf{r}^{\star}_a)^{T}(P^{\star}_a)^{-1}\bar{S}^{\star\star}_{ab}(P^{\star}_b)^{-1}\mathbf{r}^{\star}_b}
    {\displaystyle\sum_{a<b}
        \mathrm{tr}\!\left[(P^{\star}_a)^{-1}\bar{S}^{\star\star}_{ab}(P^{\star}_b)^{-1}\bar{S}^{\star\star}_{ba}\right]} \; .
    \label{eq:A2-RAD}
\end{equation}
The joint construction thus yields the astrometry-only optimal statistic for free. Consistency among $\hat A^{2}_{\rm RAD}$, $\hat A^{2}_{\rm PTA}$, and $\hat A^{2}_{\rm RAD\times PTA}$ provides a non-trivial test of the hypothesis that all three pair classes are driven by a common GWB amplitude.

\section{Scaling Laws}
\label{sec:scaling}

To provide intuition, we now derive how the SNR scales with various survey parameters and noise regimes.
As in previous work~\cite{siemens+2013, Chamberlin:2014ria}, we consider the simplified case in which the per-object spectral densities of Sec.~\ref{sec:cov_matrix} contain only the GWB and white noise (assumed the same for every object of a given type),
\begin{align}
    \mathcal{P}^{p}(f) &= \mathcal{P}^{pp}_{\rm GW}(f) + \mathcal{P}^{p}_{\rm WN} \; , \\
    \mathcal{P}^{\star}(f) &= \mathcal{P}^{\star\star}_{\rm GW}(f) + \mathcal{P}^{\star}_{\rm WN} \; ,
\end{align}
with white-noise levels set by the per-epoch precision ($\sigma_p$ for pulsars and $\sigma_\star$ for stars), and cadence of each survey ($\Delta t_p$ for PTAs and $\Delta t_\star$ for RAD),
\begin{equation}
    \mathcal{P}^{p}_{\rm WN} = 2\sigma_p^2\,\Delta t_p \; , \qquad
    \mathcal{P}^{\star}_{\rm WN} = 2\sigma_\star^2\,\Delta t_\star \; .
    \label{eq:wn_levels}
\end{equation}
The GWB power-law spectral densities are given in Eqs.~\eqref{eq:P_pp}--\eqref{eq:P_ps}.
Note that $(\mathcal{P}^{p\star}_{\rm GW})^2 = \mathcal{P}^{pp}_{\rm GW}\,\mathcal{P}^{\star\star}_{\rm GW}$, which is the combination that enters the cross-correlation SNR of Eq.~\eqref{eq:rho-cross-v2}.

The angular dependence enters only through the root-sum-squared normalized ORFs over the pairs in each class of Eq.~\eqref{eq:pair-partition}, for which we introduce the shorthand
\begin{align}
    \Xi^{pp} \equiv \sqrt{\sum_{\substack{a<b\\\mathrm{pulsars}}} \bigl(\xi^{pp}_{ab}\bigr)^2} \; &, \qquad
    \Xi^{\star\star} \equiv \sqrt{\sum_{\substack{a<b\\\mathrm{stars}}} \sum_{i,j} \bigl(\xi^{\star\star,ij}_{ab}\bigr)^2} \; , \nonumber \\
    \Xi^{p\star} \equiv& \sqrt{\sum_{\substack{a\in\mathrm{stars}\\b\in\mathrm{pulsars}}} \sum_{i} \bigl(\xi^{p\star,i}_{ab}\bigr)^2} \; ,
    \label{eq:orf_sums}
\end{align}
where, as in Sec.~\ref{sec:cov_matrix}, $a,b$ index the objects summed over in
each class (both pulsars in $\Xi^{pp}$, both stars in $\Xi^{\star\star}$, and
star $a$ with pulsar $b$ in $\Xi^{p\star}$), and $i,j$ label the components of the
astrometric deflection.

\subsection{The Weak Signal Regime}

In the weak-signal regime, each auto-spectrum is dominated by its white noise.
As derived in Ref.~\cite{Chamberlin:2014ria}, the PTA SNR is then
\begin{align}
    \langle \hat{\rho}_{\rm PTA} \rangle
    &\approx \Xi^{pp} \frac{\mathcal{A}_p\,T_p^{\beta}}{\mathcal{P}^{p}_{\rm WN}\sqrt{\beta - 1/2}} \; , \\
    &= \Xi^{pp}
    \frac{A_{\rm gw}^2}{12 \pi^2 f_0^{2\alpha}}
    \frac{T_p^{\,3-2\alpha}}{2\sigma_p^2 \Delta t_p \,\sqrt{5/2-2\alpha}} \; ,
    \label{eq:pta_ww}
\end{align}
where $T_p$ is the PTA survey baseline. 

For RAD, we must solve the integral
\begin{equation}\label{eqn:rad-full-snr}
    \langle \hat{\rho}_{\rm RAD} \rangle = \Xi^{\star\star}
    \left(2T_\star \int^{1/(2\Delta t_\star)}_{1/T_\star}
    \frac{\bigl(\mathcal{A}_\star f^{-\gamma}\bigr)^2}{\bigl(\mathcal{P}^{\star}_{\rm WN}\bigr)^2}\,df\right)^{1/2} \; ,
\end{equation}
where $T_\star$ is the RAD survey baseline. 

Performing this integral, writing the result in terms of the GWB parameters, and taking the $T_\star \gg \Delta t_\star$ limit gives
\begin{equation}
    \langle \hat{\rho}_{\rm RAD} \rangle \approx \Xi^{\star\star}
    \frac{A_{\rm gw}^2}{12 f_0^{2\alpha}}
    \frac{T_\star^{\,1-2\alpha}}{2\sigma_\star^2\,\Delta t_\star\,\sqrt{1/2-2\alpha}} \; .
    \label{eq:rad_ww}
\end{equation}
Note that PTAs and RAD scale differently with survey baseline in the weak signal regime: for $\alpha = -2/3$, $\langle \hat{\rho}_{\rm PTA} \rangle \propto T_p^{13/3}$ while $\langle \hat{\rho}_{\rm RAD} \rangle \propto T_\star^{7/3}$.

For \radxpta, the expected SNR follows from Eq.~\eqref{eq:rho-cross-v2}. With both auto-spectra white-noise dominated,
\begin{equation}
\langle\hat{\rho}_{\rm RAD\times PTA}\rangle =
\Xi^{p\star}\left(2T\int_{1/T}^{f_{\rm max}}
\frac{\mathcal{A}_p f^{-\beta}\; \mathcal{A}_\star f^{-\gamma}}{\mathcal{P}^{p}_{\rm WN}\,\mathcal{P}^{\star}_{\rm WN}}\,df\right)^{1/2},
\label{eq:cross_weak_int}
\end{equation}
where $T$ is the minimum baseline of the two data sets, $T = \min(T_p, T_\star)$, and $f_{\rm max}=1/(2\Delta t_{\rm max})$ is set by the coarser cadence, $\Delta t_{\rm max}=\max(\Delta t_p,\Delta t_\star)$.
Taking the $T\gg\Delta t_{\rm max}$ limit,
\begin{equation}
\langle\hat{\rho}_{\rm RAD\times PTA}\rangle \approx
\Xi^{p\star} \frac{\sqrt{\mathcal{A}_p\mathcal{A}_\star}\;T^{(\beta+\gamma)/2}}
{\sqrt{\mathcal{P}^{p}_{\rm WN}\,\mathcal{P}^{\star}_{\rm WN}}\,
\sqrt{(\beta+\gamma-1)/2}} \; .
\label{eq:cross_weak_ab}
\end{equation}
With $\sqrt{\mathcal{A}_p\mathcal{A}_\star}=A_{\rm gw}^2/(12\pi f_0^{2\alpha})$ and
$(\beta+\gamma-1)/2 = 3/2-2\alpha$, this becomes
\begin{equation}
\begin{split}
\langle\hat{\rho}_{\rm RAD\times PTA}\rangle &\approx
\Xi^{p\star}
\frac{A_{\rm gw}^2}{12\pi f_0^{2\alpha}} \\
&\quad\times
\frac{T^{2-2\alpha}}
{\sqrt{2\sigma_p^2\Delta t_p\;2\sigma_\star^2\Delta t_\star}\,\sqrt{3/2-2\alpha}} \; .
\end{split}
\label{eq:cross_ww}
\end{equation}
For $\alpha=-2/3$, $\langle\hat{\rho}_{\rm RAD\times PTA}\rangle\propto A_{\rm gw}^2\,T^{10/3}$.
The time exponent $(\beta+\gamma)/2$ is the average of the PTA and RAD
exponents, $13/3$ and $7/3$.

This can be rewritten in terms of the individual SNRs as
\begin{equation}
\langle\hat{\rho}_{\rm RAD\times PTA}\rangle =
c_{\beta\gamma}\,
\frac{\Xi^{p\star}}
{\bigl(\Xi^{pp}\,\Xi^{\star\star}\bigr)^{1/2}}
\sqrt{\langle\hat{\rho}_{\rm PTA}\rangle\langle\hat{\rho}_{\rm RAD}\rangle} \; ,
\label{eq:cross_weak_geo}
\end{equation}
with
\begin{equation}
c_{\beta\gamma} =
\frac{\bigl[(\beta-\tfrac12)(\gamma-\tfrac12)\bigr]^{1/4}}
{\bigl[\tfrac12(\beta-\tfrac12)+\tfrac12(\gamma-\tfrac12)\bigr]^{1/2}} \le 1 \; .
\end{equation}
The factor $c_{\beta\gamma}$ is the square root of the ratio of the geometric to the
arithmetic mean of $\beta-\tfrac12$ and $\gamma-\tfrac12$.
It equals unity only when the two spectra have the same
slope, and $c_{\beta\gamma}\simeq0.967$ for $\alpha=-2/3$.

In the weak-signal regime, the cross-correlation SNR is therefore essentially the geometric mean of the PTA and RAD SNRs, rescaled by the ratio of the angular sums.

\subsection{The Intermediate Signal Regime}

In the intermediate regime, the GWB power exceeds the per-object white noise at low frequencies.
For each tracer this defines a crossover frequency at which the two are equal,
\begin{equation}
    f^{p}_{\rm c} \equiv \left(\frac{\mathcal{A}_p}{\mathcal{P}^{p}_{\rm WN}}\right)^{1/\beta} , \qquad
    f^{\star}_{\rm c} \equiv \left(\frac{\mathcal{A}_\star}{\mathcal{P}^{\star}_{\rm WN}}\right)^{1/\gamma} ,
    \label{eq:fc}
\end{equation}
below which the integrand of the SNR saturates at $\simeq1$, and above which it falls as $(f/f_{\rm c})^{-2\beta}$ or $(f/f_{\rm c})^{-2\gamma}$, respectively.

For PTAs at large $T_p$, the SNR can be approximated as~\cite{Chamberlin:2014ria}
\begin{equation}
    \langle \hat{\rho}_{\rm PTA} \rangle \approx \Xi^{pp} \left[ 2T_p\,\kappa(\beta)\left( \frac{\mathcal{A}_p}{\mathcal{P}^{p}_{\rm WN}}\right)^{1/\beta}\right]^{1/2} \; ,
    \label{eq:pta_int}
\end{equation}
where
\begin{equation}\label{eqn:kappa}
    \kappa(\beta) = \frac{(1-\beta)\Gamma(\beta^{-1}-1) \Gamma(2-\beta^{-1}) \Gamma(1+\beta^{-1})}{\beta \Gamma(\beta^{-1})} \; ,
\end{equation}
and $\Gamma(x)$ is the Euler gamma function (not to be confused with the unnormalized ORFs of Sec.~\ref{sec:conventions}).

For RAD, the intermediate regime is $1/T_\star \ll f^{\star}_{\rm c} \ll 1/(2\Delta t_\star)$, and Eq.~\eqref{eqn:rad-full-snr} takes the same form as the PTA case,
\begin{equation}
\langle\hat{\rho}_{\rm RAD}\rangle \approx \Xi^{\star\star}
\left[\,2T_\star\,\kappa(\gamma)\left(\frac{\mathcal{A}_\star}{\mathcal{P}^{\star}_{\rm WN}}\right)^{1/\gamma}\right]^{1/2},
\label{eq:rad_int}
\end{equation}
where $\kappa$ is the same function as in Eq.~\eqref{eqn:kappa}.

In terms of the GWB parameters, both tracers scale as $\langle\hat{\rho}\rangle\propto T^{1/2}$ in the intermediate regime, with
$\langle\hat{\rho}_{\rm PTA}\rangle\propto A_{\rm gw}^{1/\beta}=A_{\rm gw}^{2/(3-2\alpha)}$ and $\langle\hat{\rho}_{\rm RAD}\rangle\propto A_{\rm gw}^{1/\gamma}=A_{\rm gw}^{2/(1-2\alpha)}$; for $\alpha=-2/3$ these are $A_{\rm gw}^{3/13}$ and $A_{\rm gw}^{3/7}$.

The weak-to-intermediate transition for either tracer occurs when the
lowest sampled frequency drops below the crossover frequency, i.e.\ for baselines exceeding $T_{\rm c}\equiv 1/f_{\rm c}$:
\begin{align}
T^{p}_{\rm c} &= \left(\frac{\mathcal{P}^{p}_{\rm WN}}{\mathcal{A}_p}\right)^{1/\beta}
 = \left(\frac{24\pi^2\sigma_p^2\Delta t_p f_0^{2\alpha}}{A_{\rm gw}^2}\right)^{1/(3-2\alpha)},
\label{eq:Tc_pta}\\
T^{\star}_{\rm c} &= \left(\frac{\mathcal{P}^{\star}_{\rm WN}}{\mathcal{A}_\star}\right)^{1/\gamma}
 = \left(\frac{24\,\sigma_\star^2\Delta t_\star f_0^{2\alpha}}{A_{\rm gw}^2}\right)^{1/(1-2\alpha)}.
\label{eq:Tc_rad}
\end{align}

For \radxpta, there are two possible intermediate regimes: PTA in the intermediate regime with RAD in the weak regime, and both in the intermediate regime.
As we show below, realistic current and future RAD surveys will always be in the weak regime.
However, we include both cases here for completeness.

For RAD in the weak regime but PTAs in the intermediate regime, we have
\begin{equation}
\langle\hat{\rho}_{\rm RAD\times PTA}\rangle \approx
\Xi^{p\star}\;
\left(2T\,\frac{\mathcal{A}_\star}{\mathcal{P}^{\star}_{\rm WN}}
\int_{1/T}^{\infty}\! df\,\frac{f^{-\gamma}}{1+(f/f^{p}_{\rm c})^{\beta}}\right)^{1/2} \; .
\end{equation}
Because $\gamma>1$, the integral is dominated by its lower limit rather
than by the PTA crossover frequency. To leading order in $(T/T^{p}_{\rm c})^{-(\gamma-1)}$,
\begin{equation}
\begin{split}
\langle\hat{\rho}_{\rm RAD\times PTA}\rangle &\approx
\Xi^{p\star}
\left[\frac{2}{\gamma-1}\,\frac{\mathcal{A}_\star}{\mathcal{P}^{\star}_{\rm WN}}\right]^{1/2} T^{\gamma/2} \\
&= \Xi^{p\star}
\frac{A_{\rm gw}}{\sqrt{-\alpha}}\,
\frac{T^{1/2-\alpha}}{\sqrt{24\,\sigma_\star^2\Delta t_\star f_0^{2\alpha}}} \; .
\end{split}
\label{eq:cross_iw}
\end{equation}
For $\alpha=-2/3$, $\langle\hat{\rho}_{\rm RAD\times PTA}\rangle\propto A_{\rm gw}\,T^{7/6}$.
Two features are notable.  First, the pulsar noise has dropped out
entirely: at the frequencies that dominate the integral the pulsars are
signal-dominated, so they act as an almost noiseless template for the
GWB against which the astrometric data are matched. Second, the SNR is
\emph{linear} in $A_{\rm gw}$, because the signal-dominated PTA
auto-power $\propto A_{\rm gw}^2$ in the denominator cancels one power of
$A_{\rm gw}^2$ from the numerator.

When both tracers are in the intermediate regime,
\begin{equation}
\begin{split}
\langle\hat{\rho}_{\rm RAD\times PTA}\rangle\! &\approx
\!\Xi^{p\star}\!\!\left[2T\int_0^\infty\!\!\!\!\!\frac{df}{\bigl[1+(f/f^{p}_{\rm c})^{\beta}\bigr]\bigl[1+(f/f^{\star}_{\rm c})^{\gamma}\bigr]}\!\right]^{\!\!\frac{1}{2}} \\
&\simeq\!
\Xi^{p\star}\!\!
\left[2T\,\kappa_1(\gamma)\left(\frac{\mathcal{A}_\star}{\mathcal{P}^{\star}_{\rm WN}}\right)^{1/\gamma}\right]^{1/2},
\end{split}
\label{eq:cross_ii}
\end{equation}
where
\begin{equation}
\kappa_1(\gamma) = \int_0^\infty\frac{dx}{1+x^\gamma} = \frac{\pi}{\gamma\sin(\pi/\gamma)} \; .
\end{equation}
The second equality of Eq.~\eqref{eq:cross_ii} holds for $f^{\star}_{\rm c}\ll f^{p}_{\rm c}$, i.e.\ when RAD is a noisier tracer than PTAs, which is the case for all sensible current and future survey configurations.
This is the RAD intermediate law, Eq.~\eqref{eq:rad_int}, with $\kappa\to\kappa_1$.
The cross-correlation then scales as $\langle\hat{\rho}_{\rm RAD\times PTA}\rangle\propto A_{\rm gw}^{1/\gamma}T^{1/2}$, the same as RAD alone in the intermediate regime.

\subsection{Results of Scaling Laws}

Here we use example PTA and RAD survey configurations to test our scaling laws and provide some numbers for intuition.
For PTA, we assume $N_p = 67$ pulsars, $\sigma_p = 100~\rm{ns}$, $\Delta t_p = 14~\rm{days}$, and $T_p = 15~\rm{years}$.
These numbers are all similar to the NANOGrav 15 year dataset \cite{NANOGrav:2023hde}.
For RAD, we assume $N_\star = 10^6$ stars, $\sigma_\star = 0.1~\rm{mas}$, $\Delta t_\star = 30~\rm{min}$, and $T_\star = 5~\rm{years}$.
The number of stars and observational baseline are consistent with Gaia \cite{Brown2021}, while the cadence is consistent with the Kepler mission \cite{Borucki2010}.
The astrometric precision is better than either mission by 1 order of magnitude.
As such, these numbers represent an optimistic future RAD survey.

\begin{figure*}
    \centering
    \includegraphics[width=0.48\linewidth]{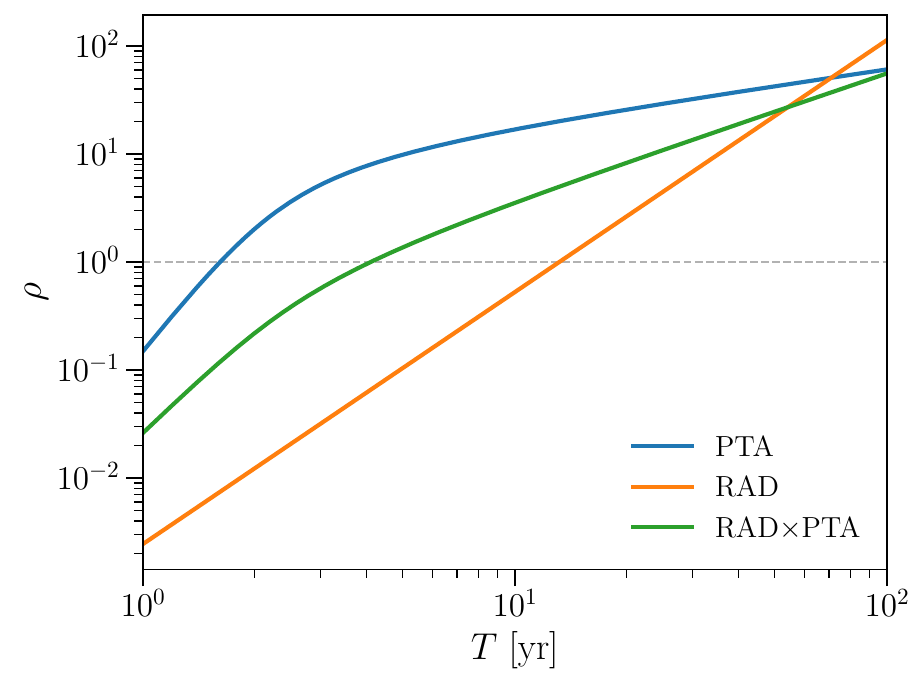}
    \includegraphics[width=0.48\linewidth]{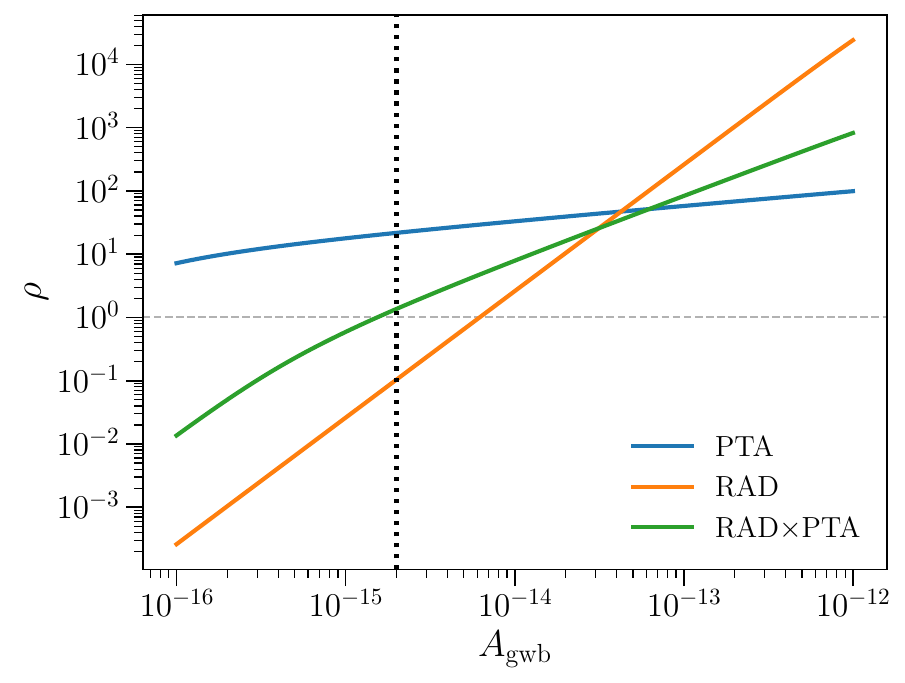}
    \caption{[\textit{Left:}] SNR vs Survey Time (in years) and [\textit{right:}] SNR vs GWB amplitude ($A_{\rm{gwb}}$) for PTAs (blue), RAD (orange), and the \radxpta cross correlation (green). We use sample survey configurations that roughly correspond to the NANOGrav 15 year dataset and an optimistic astrometric survey that combines the best features of Gaia and Kepler. 
    The PTA SNR initially scales steeply with time but quickly transitions to a shallower slope, since PTAs hit the intermediate regime quickly.
    \radxpta and RAD alone have better scaling with T because they stay in the weak regime for longer.
    Similarly, PTAs scale very slowly with $A_{\rm{gwb}}$, while \radxpta and RAD alone enjoy steeper scalings with $A_{\rm{gwb}}$, because they remain at least partially in the weak regime for all amplitudes explored here.}
    \label{fig:rho}
\end{figure*}

In the left panel of Fig.~\ref{fig:rho}, we show how the SNR scales with survey baseline, $T$, for PTAs, RAD, and \radxpta, with all baselines set equal to $T$.
The PTA curve steepens as $T^{13/3}$ below $T^{p}_{\rm c}\simeq2$~yr and
flattens toward $T^{1/2}$ above it.
RAD has a shallower weak-signal slope, $T^{7/3}$ [Eq.~\eqref{eq:rad_ww}], but keeps it across the whole figure.
For our fiducial survey ($10^6$ stars at $0.1$~mas per 30-minute exposure), $T^{\star}_{\rm c}\simeq2\times10^3$~yr, so RAD would only flatten to $T^{1/2}$ on timescales far beyond any realistic mission.
As a result, RAD overtakes the cross-correlation at $T\simeq50$~yr and the PTA at $T\simeq70$~yr.
The cross-correlation follows $T^{10/3}$ only below $T^{p}_{\rm c}$.
From $T\simeq5$~yr onward it is in the mixed regime of Eq.~\eqref{eq:cross_iw}, $\langle\hat{\rho}_{\rm RAD\times PTA}\rangle\propto A_{\rm gw}T^{7/6}$, intermediate between the two single-tracer slopes and independent of the pulsar noise.
It reaches $\langle\hat{\rho}_{\rm RAD\times PTA}\rangle=1$ at $T\simeq4$~yr, compared with $T\simeq13$~yr for RAD alone and $T\simeq1.6$~yr for the PTA.

In the right panel of Fig.~\ref{fig:rho}, we show how the SNR scales with the GWB amplitude at fixed baselines, $T_p=15$~yr and $T_\star=5$~yr; the cross-correlation uses the 5~yr overlap.
The transition amplitudes $A^{p}_{\rm c}$ and $A^{\star}_{\rm c}$ are the values of $A_{\rm gw}$ at which $T^{p}_{\rm c}=T_p$ and $T^{\star}_{\rm c}=T_\star$ in Eqs.~\eqref{eq:Tc_pta}--\eqref{eq:Tc_rad}.
For a 15~yr PTA dataset, $A^{p}_{\rm c}\simeq3\times10^{-17}$, below the plotted
range, so the PTA curve is in the intermediate regime throughout and approaches $A_{\rm gw}^{3/13}$.
RAD scales as $A_{\rm gw}^2$ across the whole figure, since $A^{\star}_{\rm c}\simeq3\times10^{-12}$ for a 5~yr baseline, just above the top of the plot.
The cross-correlation transition is set by the PTA dataset evaluated over the 5~yr overlap, $A^{p}_{\rm c}(T=5~{\rm yr})\simeq3\times10^{-16}$.
At the fiducial amplitude $A_{\rm gw}=2\times10^{-15}$ (dotted line), the PTA, cross-correlation, and RAD SNRs are $\simeq22$, $1.4$, and $0.1$.
The cross-correlation reaches $\langle\hat{\rho}_{\rm RAD\times PTA}\rangle=1$ at $A_{\rm gw}\simeq1.5\times10^{-15}$, a factor of $\simeq4$ below RAD
alone ($6\times10^{-15}$).

\section{Conclusions}
\label{sec:conclusions}

We have constructed the optimal cross-correlation statistic for a joint
astrometry--pulsar-timing dataset, extending the pulsar-timing optimal statistic
of Ref.~\cite{Chamberlin:2014ria} to a data vector spanning both stellar
deflections and timing residuals. The joint covariance organizes into three
overlap-reduction-function classes, star--star, pulsar--pulsar, and the mixed
astrometry--timing cross, and yields a single unbiased estimator of the
background amplitude, $\hat A^2_{\rm joint}$, with a clean per-channel
decomposition in which the astrometry-only statistic $\hat A^2_{\rm RAD}$ appears for free. This is one of the first optimal detection statistics to
combine datasets from two distinct classes of gravitational-wave detector. The
mixed channel is especially attractive. Since stars and pulsars have
independent noise, it is free of autocorrelations by construction, making it a
clean detection channel and a direct diagnostic of shared systematics.

The scaling laws of Sec.~\ref{sec:scaling} show that this cross channel could be more than just a consistency check. In the regime relevant to realistic surveys, the
pulsars are signal-dominated at low frequencies and act as a nearly
noiseless template for the background, so the cross-correlation SNR grows
\emph{linearly} in $A_{\rm gw}$ and crosses the detection threshold earlier than
astrometry alone. A numerical implementation and application to data are left
to future work, as is the overlap-function-dependent machinery, e.g., cosmic variance, pair covariance, and the associated null distributions, outlined in the Appendix. As astrometric surveys mature alongside
pulsar-timing arrays, the joint statistic developed here provides a natural
starting point for a first combined detection.

\begin{acknowledgments}
JSH acknowledges support from NSF CAREER Award No.~2339728 and NSF Physics Frontiers Center award No.~2020265 and No.~2607948.
JSH is also supported through an Oregon State University start-up fund.
KB and JSH thank USC for the hospitality.
A special thanks to the Aspen Center for Physics (supported by National Science Foundation grant PHY-2210452), which saw the inception and further work on this result during the 2025 winter conference on ``The Era of Binary Supermassive Black Holes: Coordination of Nanohertz-Frequency Gravitational-Wave Follow-up'' and the 2025 summer workshop on ``Physics in the Era of Nanohertz Gravitational Waves.''
\end{acknowledgments}

\appendix

\section{Supplements to the Optimal Statistic}
\label{sec:supplements}

The joint \radxpta{} statistic derived in Sec.~\ref{sec:cross-optimal} shares the mathematical form of the
pulsar-timing optimal statistic of Refs.~\cite{anholm+2009os,
Chamberlin:2014ria}: 
the estimator numerator $\hat N$ in Eq.~\eqref{eq:lnL} is a
noise-weighted quadratic form $\hat N = \sum_{ab}\mathbf{s}_a^{T}P_a^{-1}
\bar{S}_{ab}P_b^{-1}\mathbf{s}_b$, differing from the PTA case only in that
the object set now spans stars and pulsars and the correlated piece
$\bar{S}_{ab}$ carries one of three overlap reduction functions (ORFs)
rather than one. Because of this shared structure, the large body of optimal-statistic
methodology developed in the decade since Ref.~\cite{Chamberlin:2014ria} carries
over to the \radxpta{} and RAD statistics. Here we point out the connections and
leave detailed derivations to future work. It is useful to group the
developments by how they interact with our estimator: those that act only on the
quadratic-form structure and are agnostic to the ORF (Sec.~\ref{sec:quad_structure}); those whose
content lives in the specific overlap functions and so must be recomputed for
the astrometric and mixed responses (Sec.~\ref{sec:orf_methods}); and those that reconsider which
statistic one should detect with in the first place (Sec.~\ref{sec:robust_stats}).

\subsection{Quadratic structure methods}\label{sec:quad_structure}

A first family of methods operates purely on the noise-weighting and
quadratic-form structure of $\hat N$ and $\mathcal N$, independent of which ORF
enters $\bar{S}_{ab}$. These carry over verbatim to the RAD, cross, and
joint statistics.

Both RAD and PTA analyses will estimate the intrinsic noise of each object
through full Bayesian searches, fit contemporaneously with the GWB because the
low-frequency (``red'') noise and the background can be covariant. In pulsars
these processes arise from mis-modeled chromatic propagation or rotational
irregularities~\cite{ng15detchar}; in astrometry they arise from telescope and
spacecraft systematics such as attitude and reference-frame reconstruction (Stuart, et al., In Prep). In order to propagate this uncertainty into the frequentist statistic,
Ref.~\cite{vite18} introduced the noise-marginalized optimal statistic, in which
the posterior distribution of the noise parameters induces a distribution for the
statistic. Because the marginalization acts on the per-object covariances
$P_a$, it applies unchanged to $\hat A^2_{\rm joint}$ and to each
per-channel estimator in Eq.~\eqref{eq:per-channel-estimators}, now marginalizing
over both pulsar and astrometric noise posteriors.

The null distribution of the statistic---needed to convert an observed value into
a detection significance---can be built either empirically or analytically, and
both routes carry over. The earliest approach is empirical resampling: sky
scrambles~\cite{cornish+2016}, which replace the true object positions with random
ones to null the angular correlation, and phase shifts~\cite{taylor+2017}, which
randomize the phases of each object's basis functions to destroy inter-object
coherence while preserving single-object statistics. Phase shifts act per object
and are agnostic to the ORF, so they apply directly to stars, pulsars, and the
cross terms; sky scrambles instead depend on the overlap functions and must be
evaluated with the vector astrometric and mixed responses~\cite{Qin:2018yhy}
rather than the scalar timing response, so a scramble acts on both the sky
positions and the associated basis vectors (cf.\ the ORF-dependent methods
below). More recently these resampling nulls have been supplemented by analytic
distributions. As a quadratic form in approximately Gaussian residuals, both the
joint statistic and each per-channel statistic follow a generalized $\chi^2$
distribution, whose null and signal distributions and $p$-values can be computed
directly rather than by simulation~\cite{hazboun+2023gx2, vanhaasteren2025a}. This
holds without assuming $\hat N$ itself is Gaussian and provides the reference
distribution against which a \radxpta{} detection would be assessed.

Two more recent generalizations are particularly well matched to the cross
problem. The multiple-component optimal statistic~\cite{sardesai+2023} fits
several spatial correlations simultaneously rather than assuming a single ORF; in
the \radxpta{} setting this is the natural tool to fit the star--star,
pulsar--pulsar, and mixed correlations together, and to guard against a
monopolar clock error or dipolar frame error being absorbed into an apparent GWB
signal. The per-frequency optimal statistic~\cite{gersbach+2025} drops the
power-law spectral assumption and estimates the background spectrum
frequency-by-frequency; this is valuable across experiments with different
spectral reach. PTAs are most sensitive at the lowest frequencies, while RAD is
white-noise limited and bounded above only by the survey cadence,
$f_{\max}=1/2\Delta t$ , so a per-frequency cross statistic provides a direct test
that the two experiments respond to a common spectrum.

\subsection{ORF-dependent methods}\label{sec:orf_methods}

A second family depends explicitly on the overlap functions, and so carries over
in form but requires recomputing the astrometric ORF and the mixed
star--pulsar ORF~\cite{Qin:2018yhy}, together with their variances. (The sky
scrambles noted above are one instance, requiring the vector astrometric and
mixed responses.) The intrinsic
(cosmic) variance of the correlation, arising because a single realization of the
source population and a finite set of objects at fixed sky positions are
observed~\cite{allen2022, allen+2022}, becomes a set of new quantities: the
astrometric ORF and the cross ORF each carry their own cosmic variance. In
practice RAD is expected to remain measurement-noise dominated (weak-signal) for
the foreseeable future, so cosmic variance is subdominant in the RAD channel
while remaining relevant for the PTA channel. Correctly accounting for the
covariance between distinct object pairs---the pair-covariant
statistic~\cite{allen+2025}---similarly generalizes: the pair-covariance matrix
becomes block-structured across the star--star, pulsar--pulsar, and star--pulsar
classes and their cross-covariances, and is required for accurate error bars once
any channel leaves the deep weak-signal limit.

\subsection{Optimal and robust detection statistics}\label{sec:robust_stats}

A third line of work reconsiders which statistic to detect with, rather than how
to weight or calibrate the OS. The optimal statistic maximizes the deflection
(the expected signal-to-noise ratio) but does not maximize detection probability
at fixed false-alarm probability, and so is not Neyman--Pearson optimal for the
assumed signal and noise model~\cite{vanhaasteren2025b}. The Neyman--Pearson
statistic is a different quadratic form that includes autocorrelations, making it
sensitive to errors in the noise model; Ref.~\cite{vanhaasteren2025b} therefore
constructs the closest cross-correlation-only approximation to it. This
development is especially natural for \radxpta{}: the star--pulsar cross channel
is free of autocorrelations by construction, since stars and pulsars have
independent noise, so restricting to cross-correlations costs nothing there and
reinforces the role of the cross channel as a clean detection and systematics
diagnostic. The associated false-alarm thresholds can be set with the same
generalized-$\chi^2$ $p$-value machinery noted above~\cite{vanhaasteren2025a}.
Relatedly, the hybrid frequentist--Bayesian scheme can be cast as a posterior
predictive check, with a Bayesian signal-to-noise ratio as the summary
statistic~\cite{vallisneri+2023ppc}; this interpretation carries over directly to
the joint statistic.

\end{document}